# Bound State Solutions of the Dirac-Shifted Tietz-Wei Potential Plus a Generalized Ring-Shaped Potential with Spin and Pseudospin symmetry

K. O. Suleman[1,3*], K. J. Oyewumi[2+], L. A. Sunmonu[3] and D. A. Ajadi[3]

[1]Department of Physics, Nigeria Maritime University, Okerenkoko, Nigeria.

[2]Department of Physics, University of Ilorin, Ilorin, Nigeria.

[3]Department of Pure & Applied Physics, Ladoke Akintola University of Technology, Ogbomoso, Nigeria

## Abstract

In this study, approximate bound state solutions of the Dirac equation with the newly proposed shifted Tietz-Wei (sTW) potential were obtained for any arbitrary quantum number. Using Generalized Parametric Nikiforov Methods, the eigenenergy equations as well as the upper and lower spinors of the wave function corresponding to spin and pseudospin symmetric solutions were obtained by solving the radial equation. The Pekeris approximation scheme in terms of the parameters of the shifted Tietz-Wei potential was used to deal with the spin-orbit coupling potential term $\frac{k(k+1)}{r^2}$. The solutions obtained for the radial and polar angular parts of the wave functions were written in terms of the well-known Jacobi polynomials.

## 1. INTRODUCTION

Studies have shown that symmetry plays a fundamental role in Physics. A hidden symmetry in atomic nuclei, the so-called Pseudospin Symmetry (PSS), is indicated via the quasi-degeneracy that exists between single-particle. Since the discovery and introduction of the concept of PSS in atomic nuclei, comprehensive efforts have been made to understand its origin. More than a decade ago, its origin was related to a symmetry of the Dirac equation with scalar $S(\vec{r})$ and vector $V(\vec{r})$ mean-field potentials (Ginocchio, 1997). As a result, nuclear shell structure investigations has made the studying spin and pseudospin symmetric solutions of the Dirac equation an important area of research in nuclear and mathematical physics.

*kosuleman@lautech.edu.ng

+kjoyewumi66@unilorin.edu.ng

In recent years, using various methods like the traditional method, Supersymmetry (SUSY) quantum mechanics, the Nikiforov–Uvarov (NU) method, Asymptotic Iteration Method (AIM) and other approaches, considerable efforts have been made towards obtaining the bound state solutions of the Dirac equation with different potential of physical interest (Yahya, *et. al.*, 2013; Ikhdair, 2009; Ikhdair and Hamzavi, 2013; Ikhdair and Hamzavi, 2012; Oyewumi, 2012, Oyewumi and Akoshile, 2010, Setare and Haidari, 2009). The work presented here, investigates the Dirac equation within the framework of the Parametric Nikiforov-Uvarov method (Tezcan and Sever, 2009).

The newly proposed shifted Tietz-Wei (sTW) potential have been suggested and investigated under the non-relativistic limit (Falaye *et al*., 2015). The need to study this potential arises from the fact that it has been reported to account for the behaviour of many physical systems, some of which include excitons, quantum wires, quantum dots (Ikhdair and Hamzavi, 2012; Ikhdair, 2009; Khordad and Mirhosseini, 2015; Khordad and Mirhosseini, 2014).

Many researchers have in recent years applied the Nikiforov-Uvarov Method for solving both the relativistic and non-relativistic quantum mechanical problems. The method was proposed by Nikiforov and Uvarov in 1988 to solve second order differential equations with an appropriate coordinate transformation $s = (s)$.

The specific objective of this work is to obtain the bound state analytical solutions of the Dirac equation in the limit of spin and pseudospin symmetries for the shifted Tietz-Wei (sTW) potential with a Ring-Shaped potential via the parametric Nikiforov and Uvarov method.

The non-central potential of the form $V(\vec{r}) = V_{sTW}(\vec{r}) + \dfrac{1}{r^2} V_{RS}(\theta)$ studied in this work is written as:

$$V_{sTW}(\vec{r}) = D_e \left[ \frac{2(c_h - 1)e^{-b_h(r - r_e)} - (c_h{}^2 - 1)e^{-2b_h(r - r_e)}}{\left(1 - c_h e^{-2b_h(r - r_e)}\right)^2} \right] \quad V_{RS}(\theta) = \frac{\rho_1 + \rho_2 \cos^2 \theta}{\sin^2 \theta}, \qquad (1)$$

where $b_h = \beta(1 - C_h)$, $r_e$ is the molecular bond length, $\beta$ is the Morse constant, $D_e$ is the potential well depth and $C_h$ is an optimization parameter obtained from ab initio and Rydberg-Klein-Rees (RKR) intermolecular potentials, r is the intermolecular distance. Shifted by the dissociation energy $D_e$, this potential is just the Tietz-Wei (TW) potential.

This work is organized as follow: Section 2 briefly presents the method and Section 3 presents the bound state solutions of the Dirac-sTW problem in the limit of spin and pseudospin symmetries. In Section 4, a special case of the solution was obtained while Section 5 concludes the work.

## 2. Method of Solution

As mentioned earlier, an efficient mathematical method for solving quantum mechanical problems is the Nikiforov and Uvarov (NU) method. For any potential of interest, the idea of the method is to write the Schrödinger-like equation in a more general form which can be applied to any potential as follows (Tezcan and Sever, 2009):

$$\Psi''{}_n(s) + \left( \frac{c_1 - c_2 s}{s(1 - c_3 s)} \right) \Psi'{}_n(s) + \left( \frac{-\xi_1 s^2 + \xi_2 s - \xi_3}{s^2(1 - c_3 s)^2} \right) \Psi_n(s) = 0, \qquad (2)$$

A more detailed description of the approach is itemised in appendix A.

## 3. Dirac Equation for the Non-Central Shifted Tietz-Wei Potential plus a Generalized Ring Shaped Potential

The Dirac equation for any massive spin$-\frac{1}{2}$ particles interacting with arbitrary scalar potential $S(\vec{r})$, and the time-component $V(\vec{r})$ of a four-vector potential can be expressed as (Asgarifar and Goudarzi, 2013; Oyewumi, 2012; Wei and Dong, 2009; Ginocchio, 2005; Lalazissis *et al.*, 1998):

$$[c\vec{\alpha} \cdot \vec{p} + \beta(M + S(\vec{r})) + V(\vec{r})\Psi_{nk}] = E_{nk}\Psi_{nk}(\vec{r}), \Psi_{nk}(\vec{r}) = \Psi_{nk}(r, \theta, \Phi), \qquad (3)$$

where $E_{nk}$ denote the relativistic energy of the system, represents the momentum operator and $M$ mass of the particle respectively. $\boldsymbol{\alpha}$ and $\boldsymbol{\beta}$ are 4×4 matrices given by

$$\vec{\boldsymbol{\alpha}} = \begin{pmatrix} 0 & \overrightarrow{\sigma_i} \\ \overrightarrow{\sigma_i} & 0 \end{pmatrix}, \qquad \boldsymbol{\beta} = \begin{pmatrix} I & 0 \\ 0 & -I \end{pmatrix}, \tag{4}$$

where $I$ is the 2×2 unitary matrix and $\vec{\boldsymbol{\sigma}}_i$ $(i = 1,2,3)$ constitutes the three-vector Pauli spin matrices.

Following the procedure stated in (Ikhdair and Hamzavi, 2012; Oyewumi, 2012; Wei and Dong, 2009; Greiner, 2000), the Dirac spinors can be written according to the radial quantum number $\boldsymbol{n}$ and spin-orbit coupling quantum $\boldsymbol{k}$ as follows:

$$\Psi_{nk}(\vec{r}) = \frac{1}{r}\begin{pmatrix} F_{nk}(\vec{r}) & Y_{jm}^l(\theta, \Phi) \\ i\, G_{nk}(\vec{r}) & Y_{jm}^{\tilde{l}}(\theta, \Phi) \end{pmatrix} = \begin{pmatrix} f_{nk(\vec{r})} \\ g_{nk(\vec{r})} \end{pmatrix}, \tag{5}$$

where $F_{nk}(\vec{r})$ and $G_{nk}(\vec{r})$ are the radial wave functions of the upper- and lower-spinor components respectively. The spherical harmonic functions coupled to the total angular momentum $j$ and its projection $m$ on $z$ axis are denoted by $Y_{jm}^l(\theta, \Phi)$ and $Y_{jm}^{\tilde{l}}(\theta, \Phi)$.

Following the method described by Oyewumi (2012), inserting equation (5) into equation (3) yielded the following coupled differential equations:

$$\left(\frac{d}{dr} + \frac{k}{r}\right) F_{nk}(\vec{r}) = (M + E_{nk} - \Delta(\vec{r})) G_{nk}(\vec{r}), \tag{6}$$

$$\left(\frac{d}{dr} - \frac{k}{r}\right) F_{nk}(\vec{r}) = (M - E_{nk} + \Sigma(\vec{r})) F_{nk}(\vec{r}), \tag{7}$$

where $\Sigma(\vec{r}) = V(\vec{r}) + S(\vec{r})$ and $\Delta(\vec{r}) = V(\vec{r}) - S(\vec{r})$ are the sum and difference potentials, respectively. Upon elimination of $F_{nk}(\vec{r})$ and $G_{nk}(\vec{r})$ in equations (6) and (7), the upper- and lower-spinor components are obtained as second-order Schrödinger-like differential equations written as:

$$\left[\frac{d^2}{dr^2} - \frac{k(k+1)}{r^2} + \frac{\frac{d\Delta(\vec{r})}{dr}\left(\frac{d}{dr} + \frac{k}{r}\right)}{M + E_{nk} - \Delta(\vec{r})} - (M + E_{nk} - \Delta(\vec{r}))(M - E_{nk} + \Sigma(\vec{r}))\right] F_{nk}(\vec{r}) = 0, \qquad (8)$$

$$\left[\frac{d^2}{dr^2} - \frac{k(k-1)}{r^2} + \frac{\frac{d\Sigma(r)}{dr}\left(\frac{d}{dr} - \frac{k}{r}\right)}{M - E_{nk} + \Sigma(r)} - [M + E_{nk} - \Delta(\vec{r})][M - E_{nk} + \Sigma(r)]\right] G_{nk}(\vec{r}) = 0. \qquad (9)$$

Here, $k(k+1) = l(l+1)$ and the relation between $k$ and $l$ is given as:

$$k = \begin{cases} -(l+1) = \left(j + \frac{1}{2}\right), & \left(s_{1/2}, p_{3/2}, etc.\right) \; j = l + \frac{1}{2} \text{ , aligned spin}(k < 0) \\ +l = +\left(j + \frac{1}{2}\right), & \left(d_{3/2}, f_{5/2}, etc.\right), j = l - \frac{1}{2} \text{ , unaligned spin } (k > 0), \end{cases} \qquad (10)$$

while $k(k-1) = \tilde{l}(\tilde{l}+1)$, the relation between $k$ and $\tilde{l}$ is given as:

$$k = \begin{cases} -\tilde{l} = -\left(j + \frac{1}{2}\right), & \left(s_{1/2}, p_{3/2}, etc.\right) \; j = \tilde{l} - \frac{1}{2} \text{ , aligned spin}(k < 0) \\ +(\tilde{l}+1) = +\left(j + \frac{1}{2}\right), & \left(d_{3/2}, f_{5/2}, etc.\right), j = \tilde{l} + \frac{1}{2} \text{ , unaligned spin } (k > 0). \end{cases} \qquad (11)$$

### 3.1.1. Dirac equation for the radial plus angle–dependent potentials (Spin Symmetry Limit)

For the non-central potential, $\Sigma(\vec{r}) = V_{sTW}(\vec{r}) + \frac{1}{r^2} V_{RS}(\theta)$, considering spin symmetry,

$\frac{d\Delta(r)}{dr} = 0$ or $\Delta(r) = C_s$ (Asgarifar and Goudarzi 2013; Aydogdu, 2009; Aydogdu and Server, 2009; Falaye and Oyewumi, 2011; Hassanabadi *et al.*, 2012a; Hassanabadi *et al.*, 2012b; Lalazissis *et al.*, 1998; Oyewumi, 2012; Oyewumi and Akoshile, 2010), Equation (3) becomes:

$$\left[-\nabla^2 - \left(M - E_{nk} - V_{sTW}(r) + \frac{V_{RS}}{r^2}\right)(C_s - M - E_{nk})\right] f_{nk}(r) = 0, \qquad (11)$$

where we have substituted equation (5). To apply the method of variable separable, we defined

$$f_{nk}(r) = \frac{F(r)H(\theta)\Phi(\phi)}{r\sin^2\theta} \qquad (12)$$

and applying the method of variable separable, the following sets of second-order differential equations were obtained:

$$\frac{d^2\Phi(\phi)}{d\phi^2} + m^2\Phi(\phi) = 0 \quad , \tag{13a}$$

$$\frac{d^2H(\theta)}{d\theta^2} + \frac{\cos\theta}{\sin\theta}H(\theta) + \left[k(k+1) - \frac{m^2}{\sin^2\theta} - \gamma\mathcal{N}_{RS}(\theta)\right] \quad , \tag{13b}$$

$$\left[\frac{d^2}{dr^2} - \frac{k(k+1)}{r^2} - \gamma\mathcal{N}_{sTW} - \beta^2\right]F_{nk}(\vec{r}) = 0 , \tag{13c}$$

where, for mathematical simplicity, we have introduced the following notations:

$$\gamma = E_{nk} + M - C_s , \qquad \beta^2 = (E_{nk} - M)(C_s - E_{nk} - M), \tag{14}$$

and $m$ and $l$ are separation constants and $l(l+1) = k(k+1)$.

### 3.1.2. Dirac Equation for the Radial with Angle–Dependent Potential (Pseudospin Symmetry Limit)

Again, we consider radial plus angular-dependent potential $\Delta(\vec{r}) = V_{sTW}(\vec{r}) + \frac{1}{r^2}V_{RS}(\theta)$. Under the condition of pseudospin symmetry limit, $\frac{d\Sigma(r)}{dr} = 0$ or $\Sigma(r) = C_{ps}$ and equation (3) becomes

$$\left[-\vec{\nabla}^2 - \left(M - E_{nk} + W(\vec{r}) + \frac{W(\theta)}{r^2}\right)(E_{nk} - M - C_{ps})\right]g_{nk}(r) = 0 \quad . \tag{15}$$

Again, in order to apply the method of variable separable, we defined:

$$g_{nk}(r) = \frac{F(r)H(\theta)\Phi(\phi)}{r\sin^2\theta} \tag{16}$$

and obtain the following second-order differential equations:

$$\frac{d^2\Phi(\phi)}{d\phi^2} + \tilde{m}^2\Phi(\phi) = 0 \quad , \tag{17a}$$

$$\frac{d^2 H(\theta)}{d\theta^2} + \frac{\cos\theta}{\sin\theta} H(\theta) + \left[ k(k-1) - \frac{\tilde{m}^2}{\sin^2\theta} - \gamma \mathcal{N}_{RS}(\theta) \right] H(\theta) = 0 \quad , \tag{17b}$$

$$\left[ \frac{d^2}{dr^2} - \frac{k(k-1)}{r^2} - \gamma \mathcal{N}_{sTW} - \beta^2 \right] g_{nk}(\vec{r}) = 0 , \tag{17c}$$

The following notations:

$$\tilde{\gamma} = E_{nk} - M - C_{ps}, \qquad \tilde{\beta}^2 = \left( E_{nk} + M \right)\left( E_{nk} - M + C_{ps} \right), \text{ for mathematical simplicity, } \tilde{m} \text{ and }$$

$\tilde{l}$ are separation constants and $\tilde{l}\left(\tilde{l}+1\right) = k(k-1)$.

### 3.2. Bound state solution of the Dirac-Ring Shaped Shifted Tietz Wei Potential

A substitution of the potential in equation (1) into equation (13) gives the following coupled differential equations of the second order:

$$\frac{d^2 \Phi(\phi)}{d\phi^2} + m^2 \Phi(\phi) = 0 \quad , \tag{18a}$$

$$\frac{d^2 H(\theta)}{d\theta^2} + \frac{\cos\theta}{\sin\theta} H(\theta) + \left[ k(k+1) - \frac{m^2}{\sin^2\theta} - \gamma \left[ \frac{\rho_1 + \rho_2 \cos^2\theta}{\sin^2\theta} \right] \right] = 0 , \tag{18b}$$

$$\left[ \frac{d^2}{dr^2} - \frac{k(k+1)}{r^2} - \gamma \mathcal{D}_e \left[ \frac{2(c_h - 1)e^{-b_h(r-r_e)} - (c_h{}^2 - 1)e^{-2b_h(r-r_e)}}{\left(1 - c_h e^{-2b_h(r-r_e)}\right)^2} \right] - \beta^2 \right] F_{nk}(\vec{r}) = 0 . \tag{18c}$$

Taking into account the boundary condition $\Phi(\phi + 2\pi) = \Phi(\phi)$, the solution of the **φ**-dependent equation (equation 18a) can be obtained immediately as:

$$\Phi_m(\phi) = \frac{1}{\sqrt{2\pi}} e^{-im\phi}, \qquad m = 0, \pm 1, \pm 2, \dots . \tag{19}$$

### 3.2.1 Solution of the Angle-Dependent Equation

The energy eigenvalues and wave functions of the angle dependent part of the Dirac equation (18b) can be obtained by making use of the change of variable $s = \cos^2\theta$. This allows for a transformation to the form:

$$\frac{d^2 H(s)}{ds^2} + \left(\frac{\frac{1}{2} - \frac{3}{2}s}{s(1-s)}\right)\frac{dH(s)}{ds} + \frac{1}{s^2(1-s)^2}\left[\frac{1}{4}\left[-\left(k(k+1) + \gamma\rho_2\right)s^2 + \left(k(k+1) - m^2 - \gamma\rho_1\right)s\right]\right] = 0. \quad (20)$$

By comparing equation (20) with equation (2), we have the following:

$$c_1 = \frac{1}{2}, \quad c_2 = \frac{3}{2}, \quad c_3 = 1, \quad \zeta_1 = \frac{1}{4}\left(k(k+1) + \gamma\rho_2\right) \qquad \zeta_2 = \frac{1}{4}\left(k(k+1) - m^2 - \gamma\rho_2\right),$$

$$c_4 = \frac{1}{4}, \qquad c_5 = -\frac{1}{4}, \qquad c_6 = \frac{1 + 4\left(k(k+1) + \gamma\rho_2\right)}{16}, \qquad c_7 = \frac{m^2 - k(k+1) + \gamma\rho_1}{16} - \frac{1}{8},$$

$$c_8 = \frac{1}{16}, \qquad c_9 = \frac{m^2 + \gamma(\rho_1 + \rho_2)}{4} \qquad c_{10} = \frac{1}{2}, \qquad c_{11} = \sqrt{m^2 + \gamma(\rho_1 + \rho_2)}, \qquad c_{12} = \frac{1}{2}, \qquad (21)$$

$$c_{13} = \frac{1}{2}\sqrt{m^2 + \gamma(\rho_1 + \rho_2)}$$

From the energy relation given in equation (A4), we obtained:

$$\frac{3}{2}n + n\sqrt{\gamma(\rho_1 + \rho_2) + m^2} + \frac{3}{2}\sqrt{\gamma(\rho_1 + \rho_2) + m^2} + n^2 + \frac{1}{4}m^2 - \frac{1}{4}(k(k+1)) + \frac{1}{4}\gamma\rho_1 = 0. \quad (22)$$

By making the substitution $\gamma = E_{nk} + M$, the implicit dependence of the Energy $E_{nk}$ on the quantum numbers for the case of exact Spin Symmetry $(C_s = 0)$ is obtained. The corresponding wave function for the angle-dependent part is obtained as follows:

From equation (A5), we have

$$\rho(s) = s^{\frac{1}{2}}(1-s)^{\sqrt{m^2 + \gamma(\rho_1 + \rho_2)}},$$
$$\Phi(s) = s^{\frac{1}{2}}(1-s)^{\frac{1}{2}\sqrt{m^2 + \gamma(\rho_1 + \rho_2)}}, \qquad (23)$$
$$y_n = P_n^{\left(\frac{1}{2}, \sqrt{m^2 + \gamma(\rho_1 + \rho_2)}\right)}(1 - 2s),$$

with the expressions in equation (15), we obtained the angular wave function as:

$$H_{nk}(s) = N_{nk}s^{\frac{1}{2}}(1-s)^{\frac{1}{2}\sqrt{m^2 + \gamma(\rho_1 + \rho_2)}}P_n^{\left(\frac{1}{2}, \sqrt{m^2 + \gamma(\rho_1 + \rho_2)}\right)}(1 - 2s). \qquad (24)$$

where $N_{nk}$ is a normalization constant.

### 3.2.3. Solution of the Radial equation

The exact solution of equation (18c) is only possible for the s-wave ($l = 0, -1$) as a result of the presence of the spin-orbit (or pseudo) centrifugal term $\frac{k(k+1)}{r^2}$. Hence, there arises the need to employ an approximation scheme to deal with the spin-orbit (or pseudo) centrifugal term. In this work, we adopted the Pekeris-type approximation (Pekeris, 1934). After performing a new approximation for the spin-orbit term as a function of the sTW potential parameters and making change of variables $x = \frac{r - r_e}{r_e} \in (-1, \infty)$, we then write equation (18c) as;

$$\left[ \frac{d^2}{dx^2} - k(k+1) \left[ d_o + d_1 \frac{e^{-\alpha x}}{1 - c_h e^{-\alpha x}} + d_2 \frac{e^{-2\alpha x}}{\left(1 - c_h e^{-\alpha x}\right)^2} \right] \right. $$
$$\left. - \gamma \mathcal{D}_e r_e^{\,2} \left[ \frac{2(c_h - 1)e^{-\alpha x} - (c_h^{\,2} - 1)e^{-2\alpha x}}{\left(1 - c_h e^{-\alpha x}\right)^2} \right] - \beta^2 r_e^{\,2} \right] F_{nk}(x) = 0 , \quad (25)$$

where $\alpha = b_h r_e$.

Proceeding further by setting a new variable $s = e^{-\alpha x}$ allows a decomposition of the spin-symmetric Dirac equation (equation 14) into the Schrödinger-type equation satisfying the upper-spinor component $F_{nk}(s)$

$$\frac{d^2 F_{nk}}{ds^2} + \frac{1 - c_h s}{s(1 - c_h s)} \frac{dF}{ds} + \frac{1 - c_h s}{s^2 \left(1 - c_h s\right)^2} \left[ \frac{-\xi_1 s^2 + \xi_2 s - \xi_3}{\alpha^2} \right] F_{nk}(s) = 0 , \quad (26)$$

where we have made use of the following:

$$c_1 = 1, \qquad\quad c_2 = c_3 = c_h$$
$$\xi_1 = \frac{1}{\alpha^2} \left( \beta^2 r_e^{\,2} c_h^{\,2} + k(k+1)[d_o c^{\,2} - d_1 c_h + d_2] - \gamma \mathcal{D}_e r_e^{\,2} (c_h^{\,2} - 1) \right)$$
$$\xi_2 = \frac{1}{\alpha^2} \left( 2\beta^2 r_e^{\,2} c_h - k(k+1)[-2 d_o c_h + d_1 c_h] - 2\gamma \mathcal{D}_e r_e^{\,2} (c_h - 1) \right) \qquad (27)$$
$$\xi_3 = \frac{1}{\alpha^2} \left( \beta^2 r_e^{\,2} - k(k+1) d_o \right)$$

Using these constants, the energy eigenvalue equation for the relativistic sTW problem can be obtained as:

$$(2n+1)\left(\sqrt{\frac{c_h^2}{4}+\frac{1}{\alpha^2}\left(k(k+1)[2d_1c_h+d_2]+\gamma\mathcal{D}_e r_e^2(c_h-1)^2\right)}+c_h\sqrt{\frac{1}{\alpha^2}\left(k(k+1)d_o+\beta^2 r_e^2\right)}\right)+$$

$$2\left(\sqrt{\left(\frac{c_h^2}{4}+\frac{1}{\alpha^2}\left(k(k+1)[2d_1c_h+d_2]+\gamma\mathcal{D}_e r_e^2(c_h-1)^2\right)\right)\left(\frac{1}{\alpha^2}\left(k(k+1)d_o+\beta^2 r_e^2\right)\right)}\right) \qquad (28)$$

$$\frac{c_h}{2}+c_h n(n-1)+\frac{k(k+1)}{\alpha^2}d_1+\frac{2\gamma\mathcal{D}_e r_e^2}{\alpha^2}(c_h-1)=0$$

The implicit dependence of the above energy equation on the parameters of the sTW potential can be readily obtained by making the substitution:

$$\gamma = E_{nk}+M-C_s, \qquad \beta^2 = (E_{nk}+M)(C_s-E_{nk}-M). \qquad (29)$$

The weight function is obtained as:

$$\rho(s)=s^{2/\alpha\sqrt{\left(k(k+1)d_0+\beta^2 r_e^2\right)}}\left(1-c_h s\right)^{2/c_h\sqrt{\frac{c_h^2}{4}+\frac{1}{\alpha^2}\left[k(k+1)d_2+\gamma\mathcal{D}_e r_e^2(c_h-1)^2\right]}}, \qquad (30a)$$

the first and second part of the wave function can be calculated as:

$$\phi(s)=s^{1/\alpha\sqrt{\left(k(k+1)d_0+\beta^2 r_e^2\right)}}\left(1-c_h s\right)^{-c_h/2\left(\sqrt{\frac{c_h^2}{4}+\frac{1}{\alpha^2}\left[k(k+1)d_2+\gamma\mathcal{D}_e r_e^2(c_h-1)^2\right]}+\frac{c_h}{\alpha}\sqrt{\left(k(k+1)d_0+\beta^2 r_e^2\right)}\right)}$$

$$y_n(s)=P_n^{\left(2/\alpha\sqrt{\left(k(k+1)d_0+\beta^2 r_e^2\right)},\ 2/c_h\sqrt{\frac{c_h^2}{4}+\frac{1}{\alpha^2}\left[k(k+1)d_2+\gamma\mathcal{D}_e r_e^2(c_h-1)^2\right]}\right)}\left(1-2c_h s\right) \qquad (30b)$$

where $P_n^{(a,b)}(1-2c_h s)$ are the orthogonal Jacobi Polynomials.

The upper-spinor component is obtained as :

$$F_{nk}=N_{nk}s^{\frac{1}{\alpha}\sqrt{k(k+1)d_o+\beta^2 r_e^2}}\left(1-c_h s\right)^{-\frac{1}{2}+\frac{1}{c_h}\sqrt{\frac{c_h^2}{4}+\frac{1}{\alpha^2}\left[k(k+1)[2d_1c_h+d_2]+\gamma\mathcal{D}_e r_e^2(c_h-1)^2\right]}}$$

$$P_n^{\left(\frac{2}{\alpha}\sqrt{k(k+1)d_o+\beta^2 r_e^2},\ \frac{2}{c_h}\sqrt{\frac{c_h^2}{4}+\frac{1}{\alpha^2}\left[k(k+1)[2d_1c_h+d_2]+\gamma\mathcal{D}_e r_e^2(c_h-1)^2\right]}\right)}\left(1-2c_h s\right) \qquad ,$$

where $N_{nk}$ is a normalization constant.

In order to avoid repetition of procedure, the parametric mappings

$$F_{nk} \leftrightarrow G_{nk}, \quad k = k-1, \quad \gamma = \tilde{\gamma} = E_{nk} - M - C_{ps}, \quad \beta^2 = \tilde{\beta}^2 = (E_{nk} + M)(M - E_{nk} + C_{ps}). \qquad (32)$$

were used to obtain the negative energy solution of equation (18c) in the exact pseudospin symmetry case directly via the spin symmetric solution as:

$$(2n+1)\left(\sqrt{\frac{c_h{}^2}{4} + \frac{1}{\alpha^2}\left(k(k+1)[2d_1 c_h + d_2] + \tilde{\gamma}D_e r_e{}^2(c_h-1)^2\right)} + c_h\sqrt{\frac{1}{\alpha^2}\left(k(k+1)d_o + \tilde{\beta}^2 r_e{}^2\right)}\right) +$$

$$2\left(\sqrt{\left(\frac{c_h{}^2}{4} + \frac{1}{\alpha^2}\left(k(k+1)[2d_1 c_h + d_2] + \tilde{\gamma}D_e r_e{}^2(c_h-1)^2\right)\right)\left(\frac{1}{\alpha^2}\left(k(k+1)d_o + \tilde{\beta}^2 r_e{}^2\right)\right)}\right) \qquad (33)$$

$$\frac{c_h}{2} + c_h n(n-1) + \frac{k(k+1)}{\alpha^2}d_1 + \frac{2\tilde{\gamma}D_e r_e{}^2}{\alpha^2}(c_h-1) = 0$$

and the lower-spinor component of the wave functions is obtained as:

$$G_{nk} = \tilde{N}_{nk} s^{\frac{1}{\alpha}\sqrt{k(k-1)d_o + \tilde{\beta}^2 r_e{}^2}}(1 - c_h s)^{-\frac{1}{2} + \frac{1}{c_h}\sqrt{\frac{c_h{}^2}{4} + \frac{1}{\alpha^2}\left[k(k-1)[2d_1 c_h + d_2] + \tilde{\gamma}D_e r_e{}^2(c_h-1)^2\right]}},$$

$$P_n^{\left(\frac{2}{\alpha}\sqrt{k(k-1)d_o + \tilde{\beta}^2 r_e{}^2}, \frac{2}{c_h}\sqrt{\frac{c_h{}^2}{4} + \frac{1}{\alpha^2}\left[k(k-1)[2d_1 c_h + d_2] + \tilde{\gamma}D_e r_e{}^2(c_h-1)^2\right]}\right)}(1 - 2c_h s) \qquad (34)$$

where $\tilde{N}_{nk}$ is a new normalization constant.

## 4.3 A Special Case

When the potential constant $c_h$ approaches zero (i.e. $c_h \to 0$), the sTw reduces to the Morse potential and we obtained for the spin and pseudospin symmetry solution respectively, the energy equation for the Dirac-Morse potential as:

$$\frac{2n+1}{\alpha}\left(\sqrt{\left(k(k+1)d_2 + \gamma D_e r_e^2\right)}\right) + \frac{2}{\alpha^2}\left(\sqrt{\left(k(k+1)d_2 + \gamma D_e r_e^2\right)\left(k(k+1)d_o + \beta^2 r_e^2\right)}\right)$$
$$+ \frac{1}{\alpha^2}\left(k(k+1)d_1 - 2\gamma D_e r_e^2\right) = 0 \tag{35a}$$

and

$$\frac{2n+1}{\alpha}\left(\sqrt{\left(k(k-1)d_2 + \tilde{\gamma} D_e r_e^2\right)}\right) + \frac{2}{\alpha^2}\left(\sqrt{\left(k(k-1)d_2 + \tilde{\gamma} D_e r_e^2\right)\left(k(k-1)d_o + \tilde{\beta}^2 r_e^2\right)}\right)$$
$$+ \frac{1}{\alpha^2}\left(k(k-1)d_1 - 2\tilde{\gamma} D_e r_e^2\right) = 0. \tag{35b}$$

Equations (35a) and (35b) are the same as those obtained by Ikhdair and Hamzavi (2012). These equations also confirm the similarity between the sTW and the Tietz-Hua Oscillator.

## 4. CONCLUSION

By studying the Dirac equation with the recently proposed sTW potential with a ring-shaped potential system under the condition of spin and pseudospin symmetries, a relativistic solution of the system was obtained within the framework of the generalised parametric NU method. Explicit un-normalized spinor wave functions and energy eigenvalues for the system were obtained.

# Appendix

## A. Parametric Nikiforov-Uvarov Method and Pekeris approximation to the centrifugal term

To make easier, the application of the NU method in solving second order Schrödinger-like differential equation, Tezcan and Sever (2009) reported that, it is required to write our differential equation in the form:

$$\Psi''_n(s) + \left(\frac{c_1 - c_2 s}{s(1 - c_3 s)}\right)\Psi'_n(s) + \left(\frac{-\xi_1 s^2 + \xi_2 s - \xi_3}{s^2(1 - c_3 s)^2}\right)\Psi_n(s) = 0, \qquad (A1)$$

where

    i.      The essential constants are

$$
\begin{aligned}
&c_4 = \frac{1}{2}(1 - c_1) && c_5 = \frac{1}{2}(c_2 - 2c_3) \\
&c_6 = c_5^{\,2} + \zeta_1 && c_7 = 2c_4 c_3 - \zeta_2 \\
&c_8 = c_4^{\,2} + \zeta_3 && c_9 = c_3(c_7 + c_3 c_8) + c_6 \\
&c_{10} = c_1 + 2c_4 + 2\sqrt{c_8} > -1, && c_{11} = 1 - c_1 - 2c_4 + \frac{2}{c_3}\sqrt{c_9} > -1, c_3 \neq 0 \\
&c_{12} = c_4 + \sqrt{c_8} > 0, && c_{13} = -c_4 + \frac{1}{c_3}\left(\sqrt{c_9} - c_5\right), c_3 \neq 0
\end{aligned}
\qquad (A2)
$$

The essential polynomial functions:

$$
\begin{aligned}
&\pi(s) = c_4 + c_5 s - \left[(\sqrt{c_9} + c_3\sqrt{c_8})s - \sqrt{c_8}\right] \\
&k = -(c_7 + 2c_3 c_8) - 2\sqrt{c_8 c_9}, \\
&\tau(s) = c_1 + 2c_4 - (c_2 - 2c_5)s - 2\left[(\sqrt{c_9} + c_3\sqrt{c_8})s - \sqrt{c_8}\right] \\
&\tau'(s) = -2c_3 - 2\left(\sqrt{c_9} + c_3\sqrt{c_8}\right) < 0.
\end{aligned}
\qquad (A3)
$$

    (i)      The energy equation:

$$c_2 n - (2n-1)c_5 + (2n+1)\left(\sqrt{c_9} + c_3\sqrt{c_8}\right) + n(n-1)c_3 + c_7 + 2c_3 c_8 + 2\sqrt{c_8 c_9} = 0 \qquad (A4)$$

(ii)     The wave functions:

$$\rho(s) = s^{c_{10}}(1 - c_3)^{c_{11}},$$

$$\Phi(s) = s^{c_{12}}(1 - c_3)^{c_{13}}, c_{12} > 0, c_{13} > 0, \qquad (A5)$$

$$y_n(s) = P_n^{(c_{10}, c_{11})}(1 - 2c_3 s), c_{10} > -1, c_{11} > -1,$$

$$\Psi_{nk}(s) = N_{nk} s^{c_{12}}(1 - c_3 s)^{c_{13}} P_n^{(c_{10}, c_{11})}(1 - 2c_3 s).$$

$P_n^{(\mu, \nu)}(x); \mu > -1, \nu > -1, and \quad x \in [-1,1]$ are Jacobi polynomials with

$$P_n^{(\alpha, \beta)}(1 - 2s) = \frac{(\alpha + 1)n}{n!} \, {}_2F_1(-n, 1 + \alpha + \beta + n; \alpha + 1; s), \qquad (A6)$$

and $N_{nk}$ is normalization constant. Also, the above wave functions can be expressed in terms of the hypergeometric function as;

$$\Psi_{nk}(s) = N_{nk} s^{c_{12}}(1 - c_3 s)^{c_{13}} \frac{(\alpha + 1)n}{n!} \, {}_2F_1(-n, 1 + \alpha + \beta + n; \alpha + 1; s). \qquad (A7)$$

$c_{12} > 0, c_{13} > 0$ and $s \in [0, 1/c_3], c_3 \neq 0.$

For most physical Potential, the Dirac equation can only be solved for the s-wave, i.e., for the angular momentum quantum number $l = 0$ only. However, for the general solution, one needs to include the Pekeris approximation to obtain analytical solutions to the Dirac equation (Pekeris 1934, Berkdemir 2006, Ikhdair 2011).

The approximation is based on the expansion of the centrifugal term in a series exponentials depending on the nuclear distance, until the second order. By construction, this approximation is valid only for lower vibrational energy states.

Here we construct a new approximation in order to deal with the spin-orbit centrifugal (or pseudo-centrifugal) coupling terms by expanding around $r = r_e$ in a series of powers of $x = (r - r_e)/r_e \in (-1, \infty)$ as

$$V_{so}(r) = \frac{\eta_k}{r^2} = \frac{\eta_k}{r_e^2(1+x)^2} = \frac{\eta_k}{r_e^2}(1 - 2x + 3x^2 - 4x^3 + ..), \quad x \ll 1, \quad \text{(A8)}$$

where $\eta_k = l(l \pm 1)$. It is sufficient enough to keep expansion terms only up to the second order. Taking the centrifugal (pseudo centrifugal) term as

$$\tilde{V}_{so}(r) = \frac{\eta_k}{r_e^2}\left[d_0 + d_1 \frac{e^{-\alpha x}}{1 - c_h e^{-\alpha x}} + d_2 \frac{e^{-2\alpha x}}{(1 - c_h e^{-\alpha x})^2}\right], \quad x \ll 1/\alpha, \quad \alpha x \ll 1 \quad \text{(A9)}$$

where $\alpha = b_h r_e$ and $d_i$ are the range parameter and the coefficients with $(i = 0,1,2)$. After making a Taylor expansion to (A8) up to the second order term $x^2$, and then comparing equal powers with (A7), we can readily determine $d_i$ parameters as a function of the specific potential parameters $b_h, c_h$ and $r_e$ as follows:

$$d_0 = 1 - \frac{1}{\alpha}(1 - c_h)(3 + c_h) + \frac{3}{\alpha^2}(1 - c_h)^2, \qquad \lim_{c_h \to 0} d_0 = 1 - \frac{3}{\alpha} + \frac{3}{\alpha^2}, \qquad \text{(A10)}$$

$$d_1 = \frac{2}{\alpha}(1 - c_h)^2(2 + c_h) - \frac{6}{\alpha^2}(1 - c_h)^3, \qquad \lim_{c_h \to 0} d_1 = \frac{4}{\alpha} - \frac{6}{\alpha^2}, \qquad \text{(A11)}$$

$$d_2 = -\frac{1}{\alpha}(1 - c_h)^3(1 + c_h) + \frac{3}{\alpha^2}(1 - c_h)^4, \qquad \lim_{c_h \to 0} d_2 = -\frac{1}{\alpha} + \frac{3}{\alpha^2}. \qquad \text{(A12)}$$